\documentclass[final,3p,times]{elsarticle}
\usepackage{amssymb,amsmath}
\usepackage{siunitx}
\usepackage{slashed}
\usepackage{booktabs}
\usepackage{hyperref}
\DeclareMathOperator{\Tr}{Tr}

\begin{document}

\begin{frontmatter}

	\title{P-wave $c\bar{c}$ meson contributions in exotic hadrons}

	\author[1,2]{Kotaro Miyake}
	\ead{miyake@hken.phys.nagoya-u.ac.jp}

	\author[1,3,2]{Yasuhiro Yamaguchi}
	\ead{yyamaguchi@tmu.ac.jp}

	\affiliation[1]{organization={Department of Physics,
				Nagoya University},
		addressline={Furo-cho, Chikusa-ku},
		city={Nagoya},
		% citysep={}, % Uncomment if no comma needed
		% % between city and postcode
		postcode={464-8602},
		state={Aichi},
		country={Japan}}

	\affiliation[3]{organization={Kobayashi-Maskawa Institute for the Origin of Particles and the Universe,
				Nagoya University},
		addressline={Furo-cho, Chikusa-ku},
		city={Nagoya},
		% citysep={}, % Uncomment if no comma needed
		% % between city and postcode
		postcode={464-8602},
		state={Aichi},
		country={Japan}}

	\affiliation[2]{organization={Department of Physics, Tokyo Metropolitan University},
		addressline={1–1–1 Minami-Osawa},
		city={Hachioji},
		postcode={192-0397},
		state={Tokyo},
		country={Japan}}

	\begin{abstract}
		The nature of the $X(3872)$ and other exotic hadrons has been a subject of extensive investigation.
		While various theoretical models have been proposed, experimental evidence suggests that the $X(3872)$ may be a mixture state of a hadronic molecule and a $c\bar{c}$ core.
		In this work, we perform a systematic study of hidden-charm tetraquark candidates $X(3860)$, $X(3872)$, and $Z(3930)$ using a coupled-channel model that incorporates both $c\bar{c}$ states and $D^{(*)}\bar{D}^{(*)}$ hadronic molecular components.
		The model parameters are fixed to reproduce the masses of the $X(3872)$ and $Z(3930)$, and the resulting framework is used to predict the mass and structure of the $0^{++}$ state associated with the $X(3860)$.
		Our results support the mixture interpretation of these exotic hadrons, exhibiting strong attractions from the transition potential between $c\bar{c}$ and $D^{(*)}\bar{D}^{(*)}$ components.
		The molecular component dominates in the $X(3872)$, while the $c\bar{c}$ component plays a more prominent role in the $X(3860)$ and $Z(3930)$.
	\end{abstract}


\end{frontmatter}

\section{Introduction}\label{sec:intro}
Exotic hadrons, which cannot be explained by the conventional quark model, have become an important topic in hadron physics.
The discovery of the $X(3872)$ by the Belle collaboration in 2003~\cite{Belle:2003nnu} marked the beginning of a new era, revealing states that challenge our understanding of hadron structure.
Since then, numerous candidates for exotic hadrons have been reported, particularly those involving charm quarks, such as the $X$, $Y$, $Z$, $T_{cc}$, and $P_c$ states~\cite{Brambilla:2019esw,Yamaguchi:2019vea,Chen:2022asf,ParticleDataGroup:2024cfk}.

The $X(3872)$ is particularly intriguing because it is located just below the $D^0\bar{D}^{*0}$ threshold, with a mass of approximately \qty{3872}{MeV}.
Its quantum numbers are determined to be $J^{PC} = 1^{++}$, and it exhibits unusual decay patterns, including comparable branching ratios to $\omega J/\psi$ and $\pi^+\pi^- J/\psi$~\cite{ParticleDataGroup:2024cfk}, indicating significant isospin violation.
This behavior is difficult to reconcile with a pure charmonium interpretation, as the mass does not fit well within the predicted charmonium spectrum.
Instead, it suggests a composite structure, possibly as a hadronic molecule or a tetraquark.

Experimental evidence points to a mixed nature for the $X(3872)$.
Prompt production in high-energy collisions and radiative decays, such as the ratio of branching fractions for $X(3872) \to J/\psi \gamma$ versus $X(3872) \to \psi(2S) \gamma$~\cite{ParticleDataGroup:2024cfk}, indicate the presence of a compact $c\bar{c}$ component.
A plausible candidate for this component is the $\chi_{c1}(2P)$ state, predicted by constituent quark models but not yet observed experimentally.
The predicted mass of $\chi_{c1}(2P)$ is around \qty{3950}{MeV}~\cite{Godfrey:1985xj,Barnes:2005pb}, several MeV above the $D^0\bar{D}^{*0}$ threshold, making it a natural partner for the $X(3872)$ in a mixture model.

In addition to the $X(3872)$, other exotic states are considered as potential spin partners.
The $X(3860)$ with $J^{PC} = 0^{++}$ and the $Z(3930)$ with $J^{PC} = 2^{++}$ are located near the $D^{(*)}\bar{D}^{(*)}$ thresholds.
These states could correspond to the $\chi_{cJ}(2P)$ ($J=0,1,2$) charmonium states, modified by coupling to hadronic channels.

In this work, we adopt the coupled-channel mixture model of Ref.~\cite{Miyake:2025ktz} to study these hidden-charm tetraquarks.
The model treats the physical states as mixtures of bare $c\bar{c}$ cores, identified with the $\chi_{cJ}(2P)$, and hadronic molecular components composed of $D^{(*)}\bar{D}^{(*)}$ and $D_s\bar{D}_s$ channels.
We use meson exchange potentials to describe the interactions between hadrons and introduce a transition potential to couple the $c\bar{c}$ and molecular sectors.
By fitting the model parameters to the observed masses of the $X(3872)$ and $Z(3930)$, we then apply the model to the $0^{++}$ state associated with the $X(3860)$.

\section{Formalism}
\subsection{Schrödinger Equation}
We model the physical $\chi_{cJ}(2P)$ states as bound states resulting from the coupling between bare $c\bar{c}$ states and hadronic molecules.
The hadronic molecular components include various $D^{(*)}\bar{D}^{(*)}$ and $D_s\bar{D}_	s$ channels, depending on the quantum numbers $J^{PC}$; the full set of channels used in this work is listed in Table~\ref{tab:channel}.
For $0^{++}$, we consider $D_s^+ D_s^- (^1S_0)$, $D^{0} \bar{D}^{0} (^1S_0)$, $D^{+} D^{-} (^1S_0)$, and their $D$-wave counterparts. For $1^{++}$ and $2^{++}$, additional channels are included, excluding those with negligible contributions.
\begin{table}[htbp]
	\centering
	\caption{Channels included in the coupled-channel model for each $J^{PC}$ state.}\label{tab:channel}
	\begin{tabular}{@{}ll@{}}
		\toprule
		Particle  & Channel                                                                                                                                                                                             \\
		\midrule
		$X(3860)$ & $D_s^+D_s^- ({}^1S_0)$, $D^{0}\bar{D}^{0}({}^1S_0)$, $D^{+}D^{-}({}^1S_0)$, $D^{0}\bar{D}^{0}({}^5D_0)$, $D^{+}D^{-}({}^5D_0)$, $\chi_{c0}(2P)$                                                     \\
		$X(3872)$ & $D^0\bar{D}^{*0}({}^3S_1)$, $D^+D^{*-}({}^3S_1)$, $D^0\bar{D}^{*0}({}^3D_1)$, $D^+D^{*-}({}^3D_1)$, $D^{*0}\bar{D}^{*0}({}^5D_1)$, $D^{*+}D^{*-}({}^5D_1)$, $\chi_{c1}(2P)$                         \\
		$Z(3930)$ & $D_s^+D_s^- ({}^1D_2)$, $D^{0}\bar{D}^{0}({}^5S_2)$, $D^{+}D^{-}({}^5S_2)$, $D^{0}\bar{D}^{0}({}^1D_2)$, $D^{+}D^{-}({}^1D_2)$, $D^{0}\bar{D}^{0}({}^5D_2)$, $D^{+}D^{-}({}^5D_2)$, $\chi_{c2}(2P)$ \\
		\bottomrule
	\end{tabular}
\end{table}

The coupled-channel Schrödinger equation is given by:
\begin{equation}
	\mathcal{H} \Psi = E \Psi,
\end{equation}
where the Hamiltonian is:
\begin{equation}
	\mathcal{H} =
	\begin{pmatrix}
		H_0 + V_\mathrm{OBE} & \mathcal{U}^\dagger                    \\
		\mathcal{U}          & (m_{\chi_{cJ}} - m_\mathrm{threshold})
	\end{pmatrix}
	.
\end{equation}
Here, $H_0$ is the kinetic energy operator of hadronic molecules, $V_\mathrm{OBE}$ is the one-boson-exchange potential for hadron-hadron interactions, and $\mathcal{U}$ is the transition potential connecting the $c\bar{c}$ and molecular sectors.
The wave function $\Psi$ is a vector in the channel space, including the bare $c\bar{c}$ state.

\subsection{Heavy Meson Chiral Lagrangian}
The $V_\mathrm{OBE}$ is derived from effective Lagrangians~\cite{Casalbuoni:1996pg} respecting heavy quark spin symmetry, chiral symmetry, and hidden local symmetry~\cite{Bando:1987br,Harada:2003jx}.
A heavy meson field $H^{(Q)}$ consisting of a heavy pseudoscalar meson field $P$ and a heavy vector meson field $P^{*\mu}$, and its conjugate field $\bar{H}^{(Q)}$, are defined as follows to satisfy the heavy quark spin symmetry:
\begin{equation}
	H^{(Q)}_a \equiv \frac{1+\slashed{v}}{2}(P^{*\mu}_a\gamma_\mu-P_a\gamma^5),\quad\bar{H}^{(Q)}_a \equiv \gamma^0(H^{(Q)}_a)^\dagger\gamma^0 = (P^{*\mu\dagger}_a\gamma_\mu+P^\dagger_a\gamma^5)\frac{1+\slashed{v}}{2},
\end{equation}
where $v^\mu$ is the four-velocity of the heavy meson, and $a$ is the light flavor index.
The interaction Lagrangians for pseudoscalar and vector meson exchanges are given by:
\begin{align}
	\mathcal{L}_{HH\pi} & = ig \Tr[H^{(Q)}_b\gamma_\mu\gamma_5 A^\mu_{ba}\bar{H}^{(Q)}_a],                                                                          \\
	\mathcal{L}_{HHV}   & = i\beta \Tr[H^{(Q)}_bv^\mu(V_\mu-\rho_\mu)_{ba}\bar{H}^{(Q)}_a] + i\lambda \Tr[H^{(Q)}_b \sigma_{\mu\nu}F^{\mu\nu}_{ba}\bar{H}^{(Q)}_a],
\end{align}
where $A^\mu$ is the axial vector current, $V_\mu$ is the vector meson current, $\rho_\mu$ is the vector meson field, and $F_{\mu\nu}(\rho)$ is the field strength tensor of $\rho_\mu$.
The coupling constants $g$ is determined from the decay width of $D^{*+}\to D^+\pi^0$ experimental data, while
$\beta$, and $\lambda$ are determined from lattice QCD calculations.
The Lagrangian of an antiheavy meson can be obtained by taking the charge conjugate.

\subsection{One-Boson-Exchange Potential}
We include exchanges of pseudoscalar mesons ($\pi$, $\eta$, $K$) and vector mesons ($\rho$, $\omega$, $K^*$, $\phi$).

For example, the pseudoscalar exchange potential for $0^{++}$ includes terms like:
\begin{equation}
	V_\pi = \frac{1}{3} (g^2 / 2f_\pi^2) [C_\pi(r;\Lambda_\pi) + T_\pi(r;\Lambda_\pi)],
\end{equation}
where $C$ and $T$ are central and tensor components.
$\Lambda_\pi$ is a cutoff parameter, and in this study, each particle exchanged has a different cutoff parameter.
The cutoff parameter $\Lambda_\mathrm{ex}=m_\mathrm{ex}+\qty{220}{MeV}\cdot\alpha$ is used according to the mass $m_\mathrm{ex}$ of the particle to be exchanged.
The dimensionless parameter $\alpha$ is a free parameter that controls the strength of the potential, and we vary it to fit the experimental data.
Vector meson exchanges are included similarly.

\subsection{Transition Potential}
The transition potential $\mathcal{U}$ allows the bare $\chi_{cJ}(2P)$ to decay into hadronic molecules. It is modeled as~\cite{Takizawa:2012hy}:
\begin{equation}
	\mathcal{U} = \int d^3x \, 2\pi f_{spin} g_{c\bar{c}} \Lambda_q^3 e^{-\Lambda_q r} Y_{lm}(\Omega) \langle \mathbf{x} | \mathrm{HM} \rangle,
\end{equation}
where $f_{spin}$ is a spin factor from Clebsch-Gordan coefficients, $g_{c\bar{c}}$ is the coupling strength, $\Lambda_q$ regulates the range, and $\langle \mathbf{x} | \mathrm{HM} \rangle$ is the molecular wave function.

\section{Numerical Results}
We solve the coupled-channel equations numerically using the Gaussian expansion method.
The bare masses of $\chi_{cJ}(2P)$ are taken from the Godfrey-Isgur model~\cite{Godfrey:1985xj,Barnes:2005pb}: 3916 MeV for $J=0$, 3953 MeV for $J=1$, and 3979 MeV for $J=2$.

The free parameters are the cutoff $\alpha$ for the form factor, the coupling $g_{c\bar{c}}$, and the range $\Lambda_q$.
The value of $\alpha$ was varied to $0.7$, $1.0$, and $1.3$, while $g_{c\bar{c}}$ and $\Lambda_q$ were fixed to reproduce the masses of $X(3872)$ and $Z(3930)$.
The obtained values are shown in Table~\ref{tab:parameters}.
\begin{table}[htbp]
	\centering
	\caption{Model parameters obtained by fitting the masses of $X(3872)$ and $Z(3930)$.}\label{tab:parameters}
	\begin{tabular}{@{}lll@{}}
		\toprule
		$\alpha$ & $g_{c\bar{c}}$ & $\Lambda_q$ (MeV) \\
		\midrule
		0.7      & 0.0448         & 2260              \\
		1.0      & 0.0427         & 3089              \\
		1.3      & 0.0409         & 4647              \\
		\bottomrule
	\end{tabular}
\end{table}

Using these parameters, we predict a $0^{++}$ bound state at approximately 3860 MeV, matching the $X(3860)$.
The mixing ratios reveal the internal structure: the $X(3872)$ is predominantly molecular (80-85\% $D^0 \bar{D}^{*0}$), while the $X(3860)$ and $Z(3930)$ are mostly $c\bar{c}$ (90-95\%).
Expectation values of the potentials indicate that the coupling between sectors is key to the mixture nature.

\section{Summary}
We have investigated hidden-charm tetraquarks using a coupled-channel mixture model, successfully reproducing the masses of $X(3860)$, $X(3872)$, and $Z(3930)$.
$X(3860)$ and $Z(3930)$ are found to be predominantly $c\bar{c}$ states with small molecular components, while $X(3872)$ is mainly a hadronic molecule with a small $c\bar{c}$ admixture.
The analysis highlights the importance of the $c\bar{c}$-hadron coupling in generating these exotic states.
Future research will involve improving the transition amplitude of this model and extending it to handle resonance states simultaneously, enabling a more comprehensive explanation of a wider range of exotic hadrons.

\section*{Acknowledgement}
This work is supported by the RCNP Collaboration Research Network program as the project number COREnet-056.
\bibliographystyle{elsarticle-num}
\bibliography{references}
\end{document}